\documentclass[aip,apl,reprint]{revtex4-1}

\usepackage{times,amsmath,amssymb,amsfonts,bbm,graphicx}
\usepackage{bm}

\newcommand{\beq}[1]{
\begin{equation}
\label{e#1} }
\newcommand{\eeq}{
\end{equation}
}

\newcommand{\be}{\begin{equation}}
\newcommand{\ee}{\end{equation}}
\newcommand{\bea}{\begin{eqnarray}}
\newcommand{\eea}{\end{eqnarray}}

\begin{document}
\title{Surface morphology and magnetic anisotropy in (Ga,Mn)As}

\author{S. Piano}
\affiliation{School of Physics and Astronomy, University of Nottingham, University Park, Nottingham,
NG7 2RD, United Kingdom.}

\author{X. Marti}
\affiliation{Department of Condensed Matter Physics, Faculty of Mathematics and Physics,
Charles University, Ke Karlovu 5, 121 16 Prague 2, Czech
Republic.}

\author {A. W. Rushforth}
\affiliation{School of Physics and Astronomy, University of Nottingham, University Park, Nottingham,
NG7 2RD, United Kingdom.}

\author{K. W. Edmonds}
\affiliation{School of Physics and Astronomy, University of Nottingham, University Park, Nottingham,
NG7 2RD, United Kingdom.}

\author{R.~P.~Campion}
\affiliation{School of Physics and Astronomy, University of Nottingham, University Park, Nottingham,
NG7 2RD, United Kingdom.}

\author{M. Wang}
\affiliation{School of Physics and Astronomy, University of Nottingham, University Park, Nottingham,
NG7 2RD, United Kingdom.}

\author{O. Caha}
\affiliation{Institute of Condensed Matter Physics, Faculty of
Science, Masaryk University, Kotl\'{a}\v{r}sk\'{a} 2, 611 37 Brno,
Czech Republic.}

\author{T. U. Sch\"{u}lli}
\affiliation{ ESRF, BP220 Grenoble, France.}

\author{V. Hol\'y}
\affiliation{Department of Condensed Matter Physics, Faculty of Mathematics and Physics,
Charles University, Ke Karlovu 5, 121 16 Prague 2, Czech
Republic.}

\author{B. L. Gallagher}
\affiliation{School of Physics and Astronomy, University of Nottingham, University Park, Nottingham,
NG7 2RD, United Kingdom.}

\begin{abstract}
Atomic Force Microscopy and Grazing incidence X-ray diffraction measurements have revealed the presence of ripples aligned along the $[1\overline{1}0]$ direction on the surface of (Ga,Mn)As layers grown on GaAs(001) substrates and buffer layers, with periodicity of about 50 nm in all samples that have been studied. These samples show the strong symmetry breaking uniaxial magnetic anisotropy normally observed in such materials. We observe a clear correlation between the amplitude of the surface ripples and the strength of the uniaxial magnetic anisotropy component suggesting that these ripples might be the source of such anisotropy.

\end{abstract}
\pacs {75.50.Pp, 75.30.Gw, 07.79.Lh, 61.05.cp}

\maketitle

The ferromagnetic semiconductor (Ga,Mn)As has been the subject of intense research due to its potential application in spintronic devices and quantum computation \cite{Ohno,Loss}. The ferromagnetic interaction between the local Mn moments is mediated by the itinerant holes giving rise to a strong sensitivity of the magnetic properties on the carrier density and lattice strain \cite{Abolfath,Zemen}. Developing an understanding of how these interactions give rise to the magnetic properties is essential in order to realise the applications of the (Ga,Mn)As compound.
The p-d Zener model gives a satisfactory explanation for several of the properties. For instance, (Ga,Mn)As grown epitaxially on GaAs experiences a compressive which gives rise to a magnetic anisotropy with the easy axis in the plane of the layer, consistent with the predictions of the model \cite{Abolfath,Zemen}. In addition, the in plane magnetic anisotropy is found ubiquitously to consist of a cubic contribution, favouring easy axes along the $[100]$ directions, and a uniaxial component favouring one of the $[110]$ directions \cite{sawickiwang,tante}.
The presence of the cubic component is consistent with the symmetry of the GaAs lattice, but the microscopic origin of the uniaxial component has not yet been explained. Several theoretical studies have invoked the concept of a uniform anisotropic strain of order $10^{-4}$  to account for the uniaxial anisotropy component \cite{Zemen,sawickiwang,strain}, but no direct experimental evidence of such a lattice distortion has been reported \cite{welp}.

In this paper we show experimental evidence, from Atomic Force Microscopy (AFM) and Grazing incidence X-ray diffraction (GID) measurements, of the presence of periodic ripples in (Ga,Mn)As layers grown on GaAs. The analysis of AFM and GID measurements have evidenced a period of these ripples of about 50 nm.  We observe a clear correlation between the magnitude of the uniaxial anisotropy and the amplitude of the surface ripples.

\begin{figure}[b]
\centering \includegraphics [width=5.8 cm]{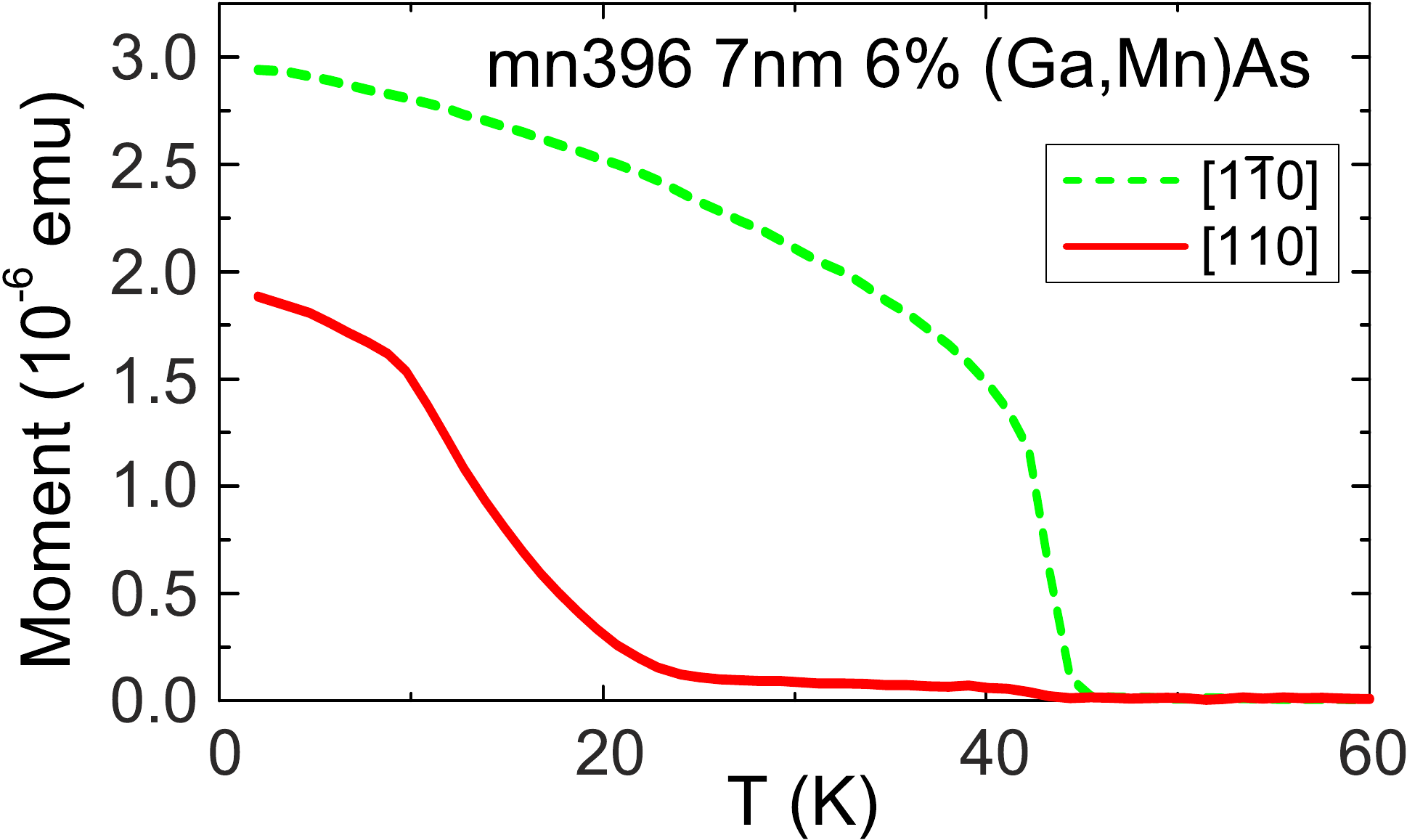}
\caption{Temperature dependence of the remnant magnetisation
along $[110]$ and $[1\overline{1}0]$ direction.}\label{magnetic}
\end{figure}

(Ga,Mn)As films with thicknesses (t),  $5 nm$, $7 nm$ and $25 nm$, and Mn concentrations, $6\%$ and $12\%$, have been deposited on GaAs (100) substrates by low temperature (${\sim}200^{\circ}$C) molecular beam epitaxy on $100$nm thick low temperature or high temperature (${\sim}580^{\circ}$C) GaAs and AlGaAs buffer layers. Atomic force microscopy (AFM) images were obtained by using an Asylum Research MFP-3D atomic force microscope. Before performing the measurements the samples were cleaned with a 1:3 HCl:H$_2$O solution to remove the superficial oxide layer. Measurements on uncleaned samples have shown the same surface structures but with additional surface contamination indicating that the cleaning was not modifying the surface morphology. The images were taken on 1x1 $\mu m^2$ areas along two different crystallographic directions, $[1\overline{1}0]$ and $[110]$. Grazing incidence X-ray diffraction (GID) measurements were performed at the beamline ID01 of the ESRF in Grenoble (France) using a photon energy of 6.5 keV and a linear detector mounted perpendicular to the sample surface. Magnetometry measurements on the as-grown films were carried out using a Quantum Design Superconducting Quantum Interference Device (SQUID) magnetometer.

\begin{figure}[t]
\centering \includegraphics [width=8.5cm]{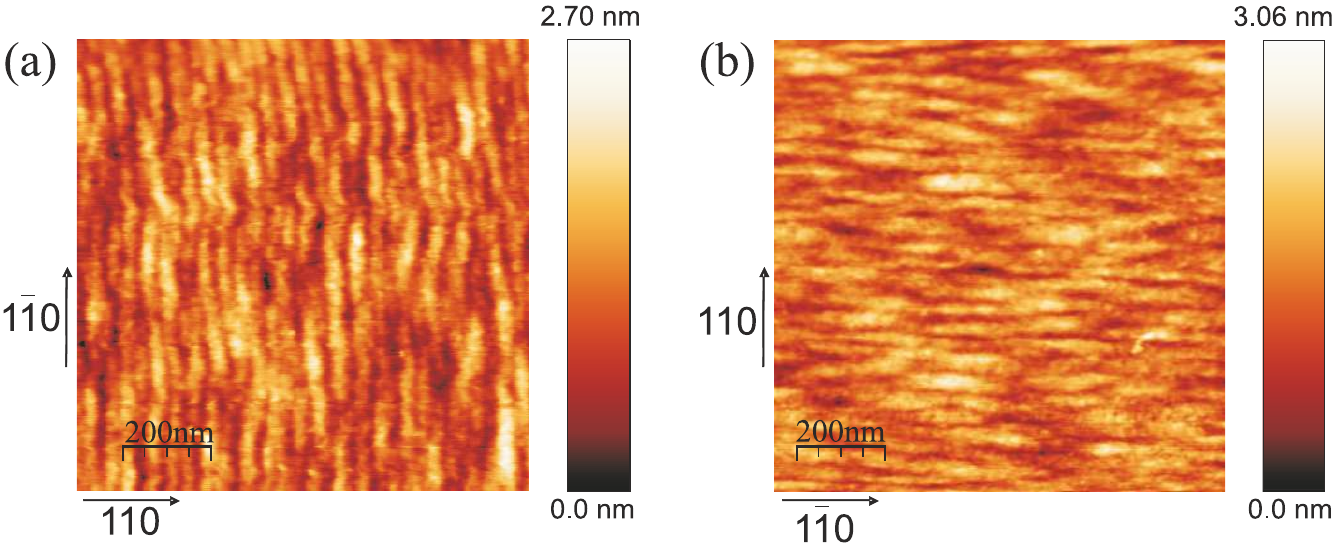}
\caption{AFM image of the sample Mn 381 a) along $[110]$
direction, and b) along  $[1\overline{1}0]$
direction.}\label{Mn381}
\end{figure}

\begin{table}[b]
  \centering
  \begin{tabular} {cccccc}
  \hline
    sample & t , Mn$\%$ & buffer & $\tau$ & RMS  & $K_u/K_c$\\
    \hline \hline
    Mn 352   & 5, 6\% & A HT & 49.7 $\pm$ 12.7 & 0.35$\pm$0.07  & 0.241 \\
    Mn 536 & 5, 6\% & B HT + B LT & 45.5 $\pm$ 11.3 & 0.31$\pm$0.04  & 0.157 \\
    \hline \hline
    Mn 396 & 7, 6\% & B HT& 47.7 $\pm$ 8.2 & 0.48$\pm$0.06 & 0.498 \\
    Mn 381 & 7, 6\% & B HT& 41.9 $\pm$ 6.4 & 0.38$\pm$0.04  & 0.325 \\
    Mn 555 & 7, 6\% & A HT&  47.7 $\pm$ 6.3 & 0.3$\pm$0.1  & 0.158 \\
    Mn 535 & 7, 6\% & B LT &  45.4 $\pm$ 10.3& 0.34$\pm$0.04 & 0.089 \\
    Mn 394 & 7, 6\% & A HT &  59.8 $\pm$ 13.5 & 0.35$\pm$0.05 & 0.249 \\
    \hline \hline
    Mn 437 & 25, 12\%  & A LT & 51.5 $\pm$ 8.6 & 0.36 $\pm$ 0.06 & 0.113 \\
    Mn 438 & 25, 12\%  & A LT & 61.7 $\pm$ 11.4 & 0.37 $\pm$0.07 & 0.149 \\
    \hline \hline
    Mn 554 & 25, 6\% & B LT &  56.0 $\pm$ 16.9 & 0.35 $\pm$ 0.04 & 0.039 \\
    Mn 499 & 25, 6\% & A HT &  47.8 $\pm$ 11.8 & 0.31$\pm$ 0.05 & 0.018 \\
    Mn 330 & 25, 6\% & A HT&  65.9 $\pm$ 11.3 & 0.31 $\pm$ 0.09  & 0.107 \\
    Mn 490 & 25, 6\% & A HT & 52.5 $\pm$ 8.3 & 0.28 $\pm$ 0.03 & -0.005 \\
           \hline
  \end{tabular}
\caption{For each sample the thickness (t) in nm, the Mn concentration (Mn \%), the buffer layer (A: GaAs; B: AlGaAs; HT: high temperature; LT: low temperature), the period ($\tau$) in nm, the RMS roughness in nm and the
magnetic anisotropy coefficient ($K_U/K_C$), are
shown.}\label{table}
\end{table}

Fig.~\ref{magnetic} shows the remnant magnetisation measured along the $[1\overline{1}0]$ and $[110]$ directions for one of our samples.
 Consistent with previous studies of unannealed (Ga,Mn)As layers we find that all samples within our study have a dominant in plane $[1\overline{1}0]$ uniaxial magnetic easy axis at high temperature, with competition between in plane  $[1\overline{1}0]$ uniaxial and in-plane $[100]$ bi-axial easy axes at low temperatures.
We consider a phenomenological description of the free energy in terms of cubic ($K_{C}$) and uniaxial ($K_{U}$) anisotropy constants: $E=K_C\sin^2(\theta)\cos^2(\theta)+K_U\sin^2(\theta)-MsH\cos(\theta-\theta_0)$, where $\theta$ and $\theta_0$ measure the direction of the magnetisation and the external magnetic field with respect to the $[110]$ crystal direction. By minimising this expression for the energy it can be shown that the relative strength of the uniaxial to cubic anisotropy
$K_U/K_C = \cos2[\tan^{-1}(M_{1\overline{1}0}/M_{110})]$, where $M_{1\overline{1}0}$ and $M_{110}$ represent the components of the magnetisation measured along $[1\overline{1}0]$ and $[110]$  respectively. We find that the temperature-dependence of the remnant magnetization after field cooling in all the studied samples is generally well-described by this phenomenological model.

Fig.~\ref{Mn381} shows AFM images along the $[1\overline{1}0]$ and $[110]$ directions for a 7nm thick (Ga,Mn)As sample containing 6\% Mn (sample Mn381). The images clearly reveal the presence of ripples aligned along the $[1\overline{1}0]$ direction. Ripple patterns such as these are known to form during the MBE growth of low temperature ($<260^{\circ}$C) GaAs in the presence of excess As. The mechanism causing the ripples to form is the suppression of diffusion of adatoms at step edges. The excess As in ref. \cite{Apostolopoulos} was believed to act as a surfactant aiding the diffusion of adatoms along the step edges. For our samples, which are not grown with a significant excess of As, we believe that Mn deposited at the surface may be playing a similar role as a surfactant. Similar ripple patterns are observed in the images obtained for all the
samples in our study. We have used Fourier analysis to extract quantitative information from these images. We can define the power spectral density as $p_s=|\phi_s|^2$, where $\phi_s$ is the Fourier amplitude corresponding to a (dimensionless) frequency $f_s=s-1$.

\begin{figure}[b]
\centering \includegraphics [width=7.5cm]{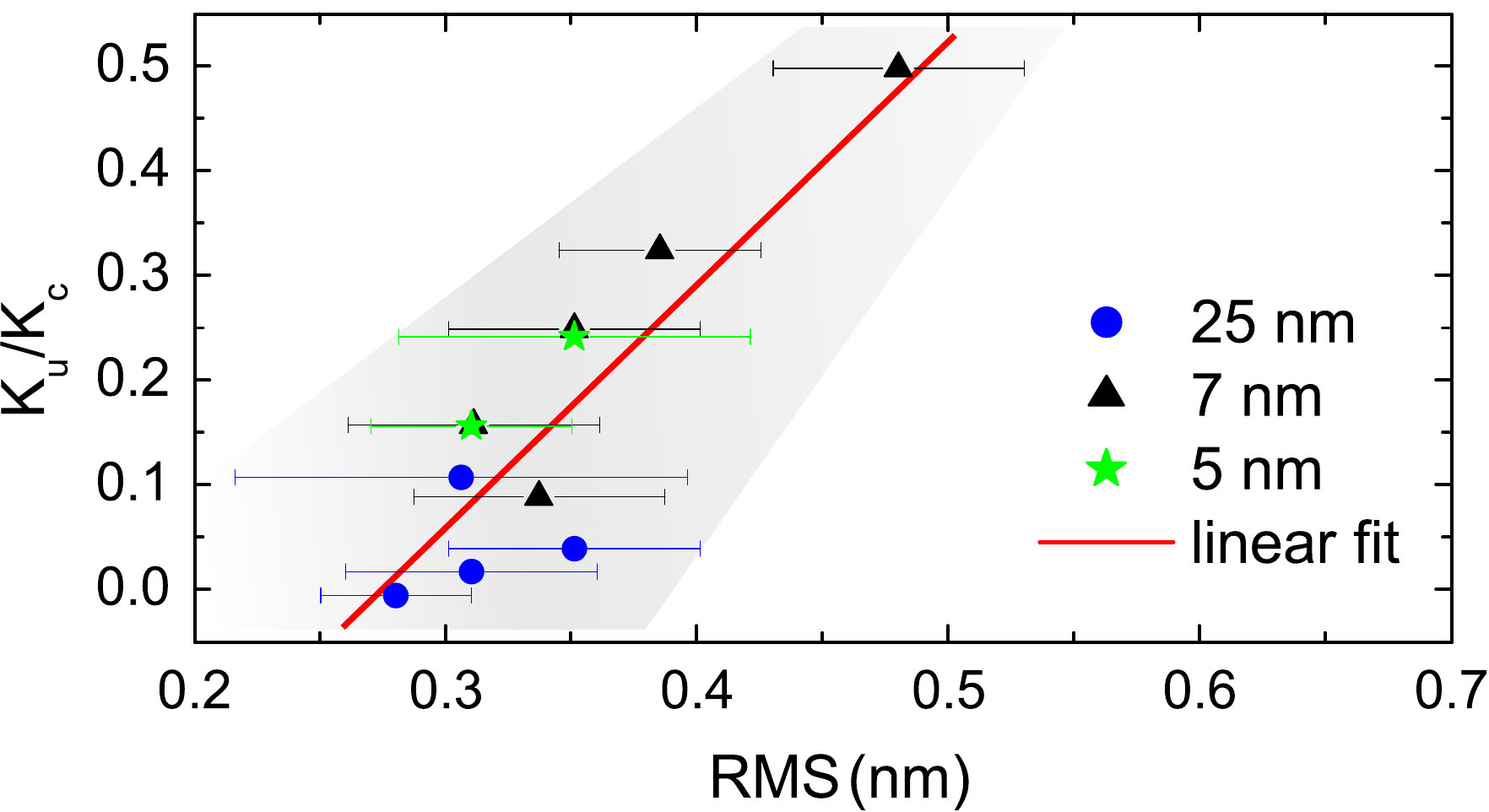} \caption{The
magnetic anisotropy coefficient, $K_U/K_C$ as a function of the
total RMS roughness for samples with $6\%$ of Mn.}\label{RMS}
\end{figure}

The mean frequency, $\overline{f}(y)$, per line is then obtained as the average of the frequencies of all Fourier harmonics (along a given scan line) weighted with the corresponding power spectral densities: $\overline{f}(y)=\big[\sum_{s=1}^{n/2}(s-1)p_s\big]/\big[\sum_{s=1}^{n/2}p_s\big]$. Then, the global `effective' mean-frequency ($\overline{f}$), characterising the oscillations in the whole image, is given by the mean $\overline{f}(y)$ of each line along all horizontal scan lines and the associated error is its standard deviation. From the $\overline{f}$ and $L$ (the width of the image) we obtain the `effective' period of the oscillations: $\tau={L}/{\overline{f}}$. Such an effective period is used to give a quantitative measure of the periodicity of the ripples observed on the AFM image.
In addition, the root mean square (RMS) roughness of each image in the direction perpendicular to the ripples is calculated by taking the standard deviation of the height for each line scan and averaging this over the whole image.

The ripples are found to be present in all of the (Ga,Mn)As samples studied. In Fig.~\ref{RMS}  we plot $K_U/K_C$ from SQUID magnetometry measurements at $2$K versus the RMS roughness, and in Table~\ref{table} we tabulate the data for each sample. In order to distinguish the effect of the roughness from effects which may alter the absolute value of $K_C$ we compare the samples within groups of similar thickness, nominal Mn concentration and Curie temperature. We find that for each group of samples there is a clear relation between the ratio of the magnetic anisotropy constants $K_U/K_C$ and the RMS roughness, in particular the influence of the uniaxial anisotropy increases as the RMS roughness increases.

\begin{figure}
\includegraphics[width=8.5 cm]{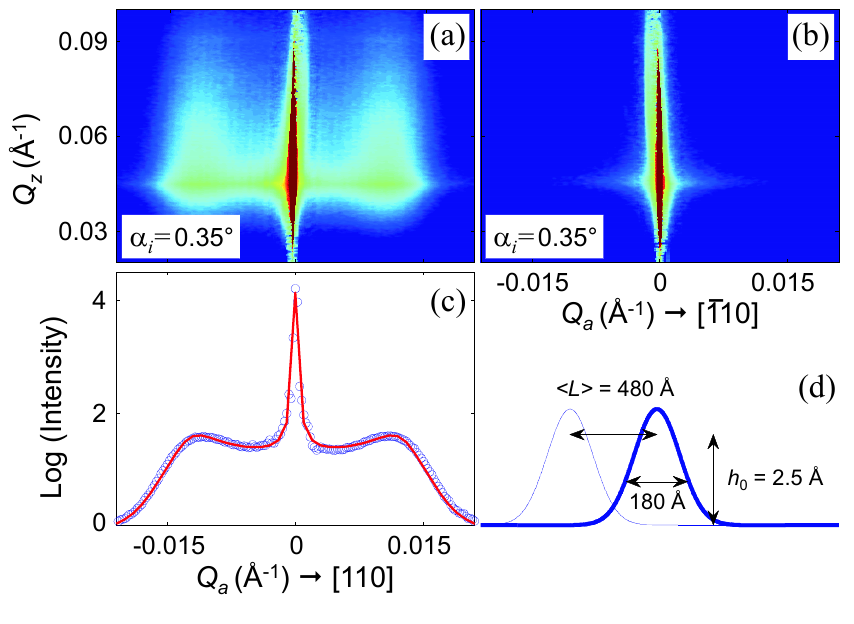}
\caption{Transverse GID scans for sample Mn381 collected along
two perpendicular directions at $\alpha_i = 0.35^{\circ}$. While
the scans with the primary beam directed across the ripples (a)
show two symmetric satellites denoting a periodicity of 48 nm,
such satellites are absent in the scans along the ripples (b). (c)
Profile from data in panel (a) and a fit assuming Gaussian-shaped
ripples are shown by circles and solid line, respectively. (d)
Sketch of the dimensions of the ripples.\label{f2}}
\end{figure}

In order to obtain information on the periodicity of the
nanostructures over the whole sample surface, grazing incidence
X-ray diffraction (GID) measurements were also performed on one sample (Mn 381). In the GID geometry, both the angle of incidence ($\alpha_i$) and the exit angle ($\alpha_f$) of the x-ray beams are close to the critical angle($\alpha_c$) of total external reflection ($0.38^{\circ}$ for the energy used) thus leading to a large footprint illuminating the sample surface, after the aperture of horizontal slits, up to a few millimeters. The difference in the azimuthal directions of the beams was set close to $2\theta_B$, where $\theta_B$ is the Bragg angle \cite{our_book}. In particular, two in-plane diffraction vectors ${\bm h} = 220$ (perpendicular to the ripples), and ${\bm h} = \bar{2}20$ (parallel to the ripples) were investigated and, for each diffraction, we measured two reciprocal-space distributions of the scattered intensity in the reciprocal planes $Q_aQ_z$ and $Q_aQ_z$, where $Q_aQ_z$ is the component of the scattering vector ${\bm Q}$ perpendicular to the surface, and $Q_a$ is the angular component of ${\bm Q}$ perpendicular to ${\bm h}$.
The angular map in Fig.~\ref{f2} (a) taken in the diffraction $\bar{2}20$ exhibits two side maxima (intensity satellites). Since $Q_a$ in this map is parallel to $[110]$, i.e. perpendicular to the ripples, these maxima stem from the periodicity of the ripples and the satellite distance is inversely proportional to the ripple period $\tau$. In turn, the width of the satellites depends on the root-mean-square deviation of the ripple spacing $\sigma_{\tau}$. From the intensity map we extracted the intensity integrated over $Q_z$ and compared it to a theoretical model. In the model, we assumed that the ripple sequence obeys the standard one-dimensional short-range order model \cite{rmp} and that the shape of a cross-section of a single ripple is Gaussian: $h(x)=h_0\exp(-x^2/R^2)$.

From Fig.~\ref{f2}(c) comparing the measured integrated intensity to the simulations it follows that the model well describes the actual ripple structure in excellent agreement with the AFM data. From the fit we determined the mean ripple period $\tau = (48\pm5) nm$, $\sigma_{\tau}=(14 \pm 3) nm$, and the shape parameters $R = (19 \pm 5) nm$ and $h_0 = (0.25 \pm5) nm$. In contrast, the angular map in the other diffraction in Fig.~\ref{f2}(b) does not show any satellites. In this map, the $Q_a$ axis is parallel to the $[\overline{1}10]$ direction, i.e. along the ripples. The surface morphology does not exhibit any pronounced space frequency in this direction so that no side maxima is observed.
From the fact that the intensity distribution along the $Q_a$ axis in Fig. \ref{f2} (a) is perfectly symmetric we can argue that the component of the strain parallel to the ripples is either zero or it does not depend on the real-space coordinate along $[110]$, i.e. perpendicular to the
ripples. A more substantial GID-based analysis on a variety of (Ga,Mn)As samples will be needed in order to possibly relate the formation of the periodic structures with the presence of uniaxial strain, that may lead to a deeper understanding of the origin of the magnetic anisotropy; this will be the subject of further work.

In conclusion, AFM and X-ray GID measurements have revealed the presence of ripples aligned along the $[1\overline{1}0]$ direction on the surface of (Ga,Mn)As layers grown on GaAs(001) substrates and buffer layers. We find that the strength of the uniaxial anisotropy of these layers increases with the amplitude of the ripples. This suggests that the symmetry breaking uniaxial anisotropy observed in (Ga,Mn)As might be related to these symmetry breaking structural features.

We thank P. Moriarty, M.  Humphry and G. Adesso for fruitful discussions. Funding from the EU (Grant No. NAMASTE-214499 and Grant No. SemiSpinNano-237375) is acknowledged.

\end{document}